\documentclass[sn-nature,Numbered]{sn-jnl}% Style for submissions to Nature Portfolio journals
\usepackage{graphicx}%
\usepackage{multirow}%
\usepackage{amsmath,amssymb,amsfonts}%
\usepackage{amsthm}%
\usepackage{mathrsfs}%
\usepackage[title]{appendix}%
\usepackage{xcolor}%
\usepackage{textcomp}%
\usepackage{manyfoot}%
\usepackage{booktabs}%
\usepackage{algorithm}%
\usepackage{algorithmicx}%
\usepackage{algpseudocode}%
\usepackage{listings}%
\theoremstyle{thmstyleone}%
\theoremstyle{thmstyletwo}%

\theoremstyle{thmstylethree}%

\begin{document}

\title[Star-Planet Interactions, Exoplanets, and Space Weather]{Radio Signatures of Star-Planet Interactions, Exoplanets, and Space Weather}


\author*[1,2]{\fnm{J.\,R.} \sur{Callingham}}\email{callingham@astron.nl}

\author[3,4]{\fnm{B.\,J.\,S.} \sur{Pope}} 

\author[1]{\fnm{R.\,D.} \sur{Kavanagh}}

\author[2]{\fnm{S.} \sur{Bellotti}} 

\author[5]{\fnm{S.} \sur{Daley-Yates}} 

\author[6]{\fnm{M.} \sur{Damasso}}

\author[7,8]{\fnm{J.-M.} \sur{Grie{\ss}meier}}

\author[9]{\fnm{M.} \sur{G\"{u}del}} 

\author[10]{\fnm{M.} \sur{G\"{u}nther}}

\author[]{\fnm{M.\,M.} \sur{Kao}$^\text{11,12}$} % weird latex bug. Have to force affliation

\author[13]{\fnm{B.} \sur{Klein}} 

\author[14,15]{\fnm{S.} \sur{Mahadevan}}

\author[16]{\fnm{J.} \sur{Morin}}

\author[17]{\fnm{J.\,D.} \sur{Nichols}}

\author[18,19]{\fnm{R.\,A.} \sur{Osten}} 

\author[20]{\fnm{M.} \sur{P\'{e}rez-Torres}} 

\author[21]{\fnm{J.\,S.} \sur{Pineda}}

\author[]{\fnm{J.} \sur{Rigney}$^\text{22,23,24}$} 

\author[25]{\fnm{J.} \sur{Saur}} 

\author[26,27]{\fnm{G.} \sur{Stef\'{a}nsson}}

\author[28,29]{\fnm{J.\,D.} \sur{Turner}}

\author[1,29]{\fnm{H. }\sur{Vedantham}}

\author[2]{\fnm{A.\,A.} \sur{Vidotto}} 

\author[30]{\fnm{J.} \sur{Villadsen}} 

\author[31,8]{\fnm{P.} \sur{Zarka}} 

\affil[1]{\orgdiv{ASTRON}, \orgname{Netherlands Institute for Radio Astronomy}, \orgaddress{\street{Oude Hoogeveensedijk 4}, \city{Dwingeloo}, \postcode{7991\,PD}, \country{The Netherlands}}}

\affil[2]{\orgdiv{Leiden Observatory}, \orgname{Leiden University}, \orgaddress{\street{PO\,Box 9513}, \postcode{2300\,RA}, \city{Leiden}, \country{The Netherlands}}}

\affil[3]{\orgdiv{School of Mathematics and Physics}, \orgname{University of Queensland}, \orgaddress{\city{St Lucia}, \postcode{QLD~4072}, \country{Australia}}}

\affil[4]{\orgdiv{Centre for Astrophysics}, \orgname{University of Southern Queensland}, \orgaddress{\street{West Street}, \city{Toowoomba}, \postcode{QLD~4350}, \country{Australia}}}

\affil[5]{School of Physics and Astronomy, University of St Andrews, North Haugh, St Andrews, Fife KY16 YSS, UK}

\affil[6]{INAF - Osservatorio Astrofisico di Torino, Via Osservatorio 20, I10025 Pino Torinese, Italy}

\affil[7]{LPC2E, OSUC, Univ Orleans, CNRS, CNES, Observatoire de Paris, F-45071 Orleans, France}

\affil[8]{\orgdiv{Observatoire Radioastronomique de Nançay (ORN)}, \orgname{Observatoire de Paris, CNRS, PSL, Université d'Orléans, OSUC}, \orgaddress{\street{route de Souesmes}, \city{Nançay}, \postcode{18330}, \country{France}}}

\affil[9]{Department of Astrophysics, University of Vienna, T\"{u}rkenschanzstr. 17, 1180, Vienna, Austria}

\affil[10]{European Space Agency (ESA), European Space Research and Technology Centre (ESTEC), Keplerlaan 1, 2201 AZ Noordwijk, The Netherlands}

\affil[11]{Department of Astronomy \& Astrophysics, University of California, Santa Cruz, CA, USA}
\affil[12]{Lowell Observatory, Flagstaff, AZ, USA}

\affil[13]{Department of Physics, University of Oxford, OX13RH, Oxford, UK}

\affil[14]{Department of Astronomy \& Astrophysics, Pennsylvania State University, University Park, PA, 16802, USA}

\affil[15]{Center for Exoplanets and Habitable Worlds, Pennsylvania State University, University Park, PA, 16802, USA}

\affil[16]{LUPM, Universit\'e de Montpellier, CNRS, Place Eug\`ene Bataillon, F-34095 Montpellier, France}

\affil[17]{Department of Physics and Astronomy, University of Leicester, Leicester, UK}

\affil[18]{\orgname{Space Telescope Science Institute}, \orgaddress{\city{Baltimore}, \postcode{MD~21218}, \country{USA}}}

\affil[19]{\orgdiv{Center for Astrophysical Sciences, Department of Physics and Astronomy}, \orgname{Johns Hopkins University}, \orgaddress{\city{3400 North Charles Street, Baltimore}, \postcode{MD~21218}, \country{USA}}}

\affil[20]{\orgdiv{Instituto de Astrofísica de Andalucía (IAA-CSIC)}, \orgname{Consejo Superior de Investigaciones Científicas (CSIC)}, \orgaddress{\street{Glorieta de la Astronomía, s/n}, \city{Granada}, \postcode{E-18008}, \country{Spain}}}

\affil[21]{Laboratory for Atmospheric and Space Physics, University of Colorado Boulder, Boulder, CO, USA}

\affil[22]{Astronomy \& Astrophysics Section, DIAS Dunsink Observatory, Dublin Institute for Advanced Studies, Dublin, D15 XR2R, Ireland}

\affil[23]{Armagh Observatory and Planetarium, College Hill, Armagh BT61 9DG, N. Ireland}

\affil[24]{School of Mathematics and Physics, Queen’s University Belfast, University Road, Belfast BT7 1NN, N. Ireland}

\affil[25]{Institute of Geophysics and Meteorology, University of Cologne, Albertus-Magnus-Platz, 50923 Cologne, Germany}

\affil[26]{Department of Astrophysical Sciences, Princeton University, 4 Ivy Lane, Princeton, NJ 08540, USA} 
\affil[27]{NASA Sagan Fellow}

\affil[28]{Department of Astronomy and Carl Sagan Institute, Cornell University, Ithaca, NY, USA}

\affil[29]{Kapteyn Astronomical Institute, University of Groningen, PO Box 72, 97200 AB, Groningen, The Netherlands}

\affil[30]{Department of Physics \& Astronomy, Bucknell University, Lewisburg, PA, USA}

\affil[31]{\orgdiv{LESIA}, \orgname{Observatoire de Paris, Université PSL, CNRS, Sorbonne Université, Université de Paris}, \orgaddress{\street{5 place Jules Janssen}, \city{Meudon}, \postcode{92195}, \country{France}}}

% \affil[31]{School of Sciences, European University Cyprus, Diogenes Street, Engomi, 1516 Nicosia, Cyprus}


\abstract{

Radio detections of stellar systems provide a window onto stellar magnetic activity and the space weather conditions of extrasolar planets, information that is difficult to attain at other wavelengths. There have been recent advances observing auroral emissions from radio-bright low-mass stars and exoplanets largely due to the maturation of low-frequency radio instruments and the plethora of wide-field radio surveys. To guide us in placing these recent results in context, we introduce the foremost local analogues for the field: Solar bursts and the aurorae found on Jupiter. We detail how radio bursts associated with stellar flares are foundational to the study of stellar coronae, and time-resolved radio dynamic spectra offers one of the best prospects of detecting and characterising coronal mass ejections from other stars. We highlight the prospects of directly detecting coherent radio emission from exoplanetary magnetospheres, and early tentative results. We bridge this discussion to the field of brown dwarf radio emission, in which their larger and stronger magnetospheres are amenable to detailed study with current instruments. Bright, coherent radio emission is also predicted from magnetic interactions between stars and close-in planets. We discuss the underlying physics of these interactions and implications of recent provisional detections for exoplanet characterisation. We conclude with an overview of outstanding questions in theory of stellar, star-planet interaction, and exoplanet radio emission, and the prospects of future facilities in answering them. 
}

% \keywords{keyword1, Keyword2, Keyword3, Keyword4}

\maketitle

We know that the vast majority of stars host planets \cite{Gaudi2021} -- now the focus of exoplanet science has shifted to understanding their detailed physics. Determining the composition of exoplanetary atmospheres, geospheres, and associated space weather conditions is considered paramount for assessing the potential habitability of a planet \cite{2019ARA&A..57..617M}. While transmission spectroscopy is beginning to reveal the composition of exoplanets \cite{wasp39jwst,Kempton2023,trappist1bjwst,trappist1cjwst}, the space weather and radiation environment that shape their atmospheric evolution and habitability remain relatively unknown.

The largest contributors to space weather in the Solar System are coronal mass ejections (CMEs) \cite{2015A&A...580A..80K,2016Ap&SS.361..253B}, which launch a significant amount of hot, dense plasma from the Sun into the Solar System. The persistent impact of CMEs on a terrestrial planet has the potential to erode its atmosphere \cite{2013oepa.book.....L,2007AsBio...7...85S}, as may the quiescent solar wind, particularly for planets without intrinsic magnetic fields \cite{kulikov07, curry15}. Despite the importance of understanding the plasma physics of CMEs, there has yet to be an unambiguous detection of a CME from a star other than our Sun. The winds of low-mass main sequence stars are also generally too tenuous to detect via current technology \cite{fichtinger17, ofionnagain19} for all but a handful of stars \cite{Wood2021}. Furthermore, there has not yet been a direct measurement of an exoplanet's magnetic field, which is crucial information for understanding its atmospheric evolution, as a planet's magnetic field could protect the atmosphere from the impact of stellar plasma \cite{owen14, vidotto20}. 

Radio observations, particularly at low frequencies ($\lesssim$300\,MHz), are a unique probe of stellar and planetary plasma environments \cite{1985ARA&A..23..169D,2019ApJ...871..214V,lazio2024}. As observed on our Sun, CMEs are often accompanied by bursty, low-frequency radio emission that encodes the kinematics of the plasma as it is ejected into interplanetary space \cite{1985ARA&A..23..169D,1985srph.book.....M}. Furthermore, the incident solar wind on the magnetised planets drives auroral emission \cite{dungey61}, particularly at radio wavelengths \cite{Zarka2007}. While optical emission from exoplanet aurorae will likely be difficult to observe even with extremely large telescopes \cite{Luger2017}, direct detection of auroral radio emission from an exoplanet has been a long-standing goal of the radio astronomy community. This would allow us to infer the presence, topology, and strength of exoplanetary magnetic fields for the first time \cite{Yantis1977,2004ApJ...612..511L,Zarka2007}. At other wavelength regimes, such information about the magnetic field of an exoplanet is either model dependent \cite{Cauley2019} or untraceable \cite{lazio2024}.

While the goals of detecting CMEs, stellar winds, and planetary magnetic fields in exoplanetary systems have been pursued for decades, recent observational progress has been made due to both the maturation of low-frequency radio arrays and the increased sensitivity at gigahertz-frequencies. For example, observations from the LOw-Frequency ARray (LOFAR) \cite{vanHaarlem2013}, Giant Metrewave Radio Telescope (GMRT), Karl G. Jansky Very Large Array (JVLA), Five-hundred-meter Aperture Spherical Telescope (FAST), and Australian Square Kilometre Array Pathfinder (ASKAP) \cite{Johnston2008} have begun to uncover new and diverse signals from radio stellar systems \cite{2024arXiv240407418D}, some potentially consistent with interactions between stars and planets analogous to some phenomena observed in our Solar System \cite{Vedantham2020, Zic2020, PerezTorres2021, callingham21, Turner2021, Pritchard2021,2023ApJ...953...65Z}. This advancement in observational radio astronomy has also been paired with a revolution in optical and near-infrared facilities, in particular the Transiting Exoplanet Survey Satellite (TESS) \cite{tess}, near-infrared radial velocity \cite{carmenes,hpf} and Zeeman-Doppler imaging (ZDI) \cite{semel89,Morin2010} surveys. The combination of these radio and optical/near-infrared facilities provides an unparalleled opportunity to trace key stellar activity indicators across the electromagnetic spectrum, informing us about whether emission is driven by coronal or magnetospheric processes, and providing probes of extrasolar space weather environments. 

This manuscript is focused on communicating the recent observational advances that have been made on coherent radio signatures of star-planet interactions, exoplanets, ultracool dwarfs, and space weather, and the associated basic foundational theories -- with an aim to be understandable and a primer for new doctoral students and scholars to the field. %We begin with a short, general description of the radio emission from bodies in the Solar System as they act as the local analogues of stellar and exoplanetary radio emission. We then summarize recent progress on radio observations of stars, exoplanets, ultracool dwarfs, and star-planet interactions. We focus both on what such observations tell us about stellar activity and extrasolar space weather conditions. We conclude this review with an outlook for the field with the Square Kilometer Array (SKA; \cite{Dewdney2009}) and Next Generation Very Large Array (ngVLA; \cite{DiFrancesco2019}) becoming operational in the next decade, and recommend ideal targets and strategies to pursue to ensure radio observations of stars and planets delivers on its potential.

%%%%%%%%%%%%%%%%%%%%%%%%%%%%%%%%%%%%%%%%%%%%%%%%

\section{Radio emission in the Solar System}
\label{sec:sun_jup}
The properties of radio emission observed from the Sun and Solar System planets are often used as templates for interpreting radio emission from stellar systems. Solar radio emission is often incoherent, implying the electrons producing the radio emission do not act in phase, and is observed over a broad range of frequencies up to a few gigahertz \cite{1985ARA&A..23..169D,bastian1998}. Such incoherent radio emission is often associated with magnetic reconnection events in the Solar corona, or active regions on the Sun, where mildly-relativistic electrons produce gyrosynchrotron emission by their motion along magnetic field lines. 

\begin{figure}
    \centering
    \includegraphics[width = \textwidth]{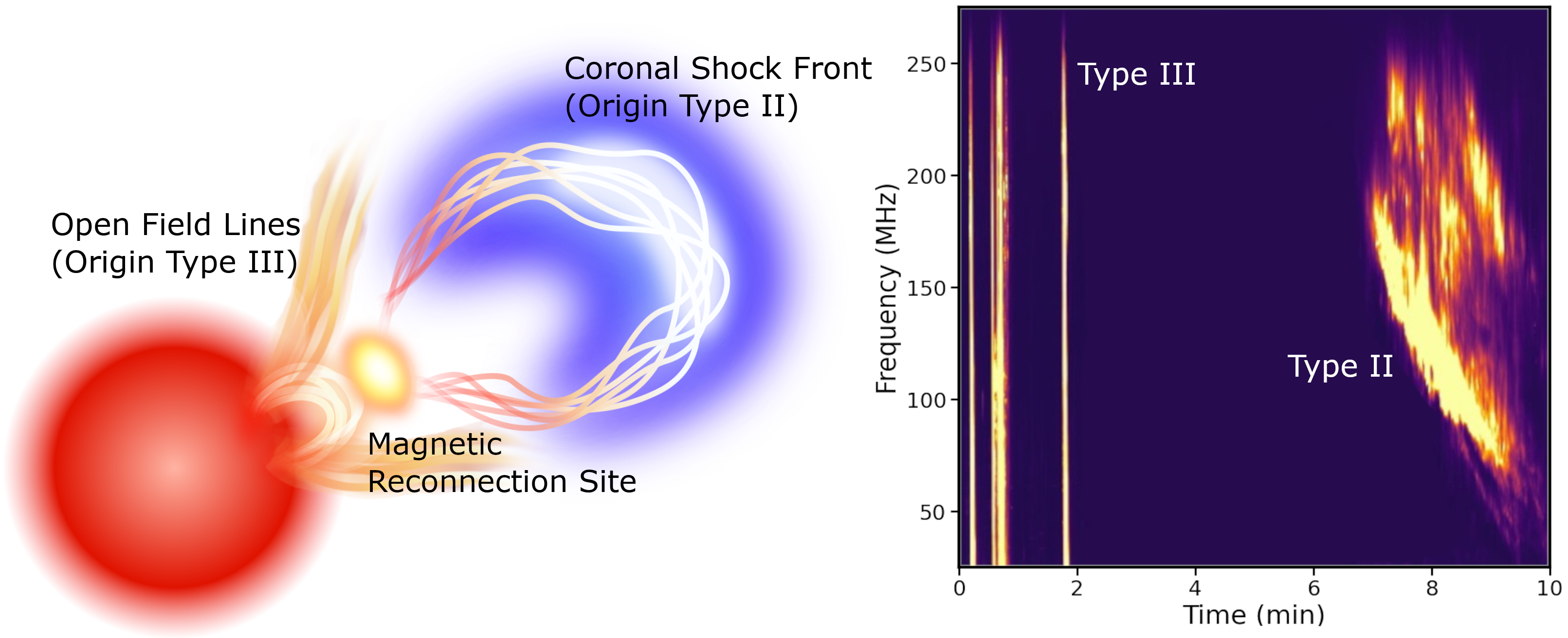}
    \caption{Schematic representation of a CME and the radio dynamic spectrum of the event that shows Type\,II and III bursts. The Type\,II burst is produced in the coronal shock front, as represented by the blue region emanating from the red star in the left panel. The magnetic reconnection event that is allowing mass to escape the magnetosphere of the star is shown as the yellow region. Type\,III bursts are produced on open field lines surrounding the magnetic reconnection event. Some structure and higher frequency harmonic emission is evident in and around the Type\,II burst in the right panel. The dynamic spectrum is in total intensity.}
    \label{fig:cme_typeii}
\end{figure}

\begin{figure}
    \centering
    \includegraphics[width = \textwidth]{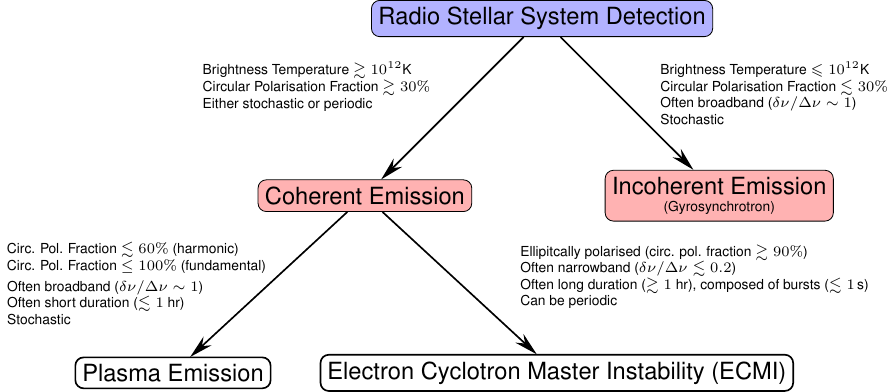}
    \caption{Schematic for distinguishing the emission mechanism operating when a radio stellar system is detected. The emission characteristics listed for differentiating between the different mechanisms should be treated an approximate guideline. For a more physical differentiation based on the conditions of the plasma, the reader is referred to  \cite{1985ARA&A..23..169D,bastian1998}. $\delta\nu/\Delta\nu$ represents the amount of bandwidth $\delta\nu$ the emission occupies relative to the available bandwidth $\Delta\nu$, assuming a relative large fractional bandwidth. Differentiating between plasma emission and emission from the electron cyclotron maser (ECM) instability can be difficult if the time-frequency structure of the radio emission can not be resolved. In that case, arguments can be made in favour of one emission mechanism based on the coronal scale height of the radio star, as derived from the stellar X-ray luminosity \cite{2015A&A...578A.129J,2020A&A...639L...7V}. Fundamental and harmonic plasma emission have different circular polarisation fractions and maximum brightness temperatures, with harmonic plasma emission able to reach the highest brightness temperatures but limited to $\lesssim 60\%$ circular polarisation fraction \cite{2020A&A...639L...7V}. Note that the polarisation fractions reported do not take into account propagation effects, which often suppresses the fractional amount \cite{Hallinan2006ApJ...653..690H}.}
    \label{fig:emission_mech}
\end{figure}

Coherent solar radio emission is predominantly generated by plasma emission via plasma waves. These plasma waves are excited by beams of energetic electrons that have been accelerated by magnetic reconnection events or shocks \cite{1985ARA&A..23..169D,bastian1998}. Phenomenologically, the two most important types of solar bursts for tracing space weather are classed as Type~II and Type~III bursts. Type~II bursts, lasting from several minutes to hours, are usually produced by the acceleration of electrons by shocks at the leading edges of outward-moving CMEs. Type~II bursts tend to slowly decrease in frequency over several minutes, providing a measure of how the plasma density decreases as the wave propagates out of the solar corona. Type~III bursts are short ($\sim$ few seconds) events associated with electron beams accelerated by reconnection events, and often accompany Type~II bursts. A schematic of the set up of a CME, and the dynamic spectrum of Type\,II and III bursts, is provided in Figure\,\ref{fig:cme_typeii}. In Figure\,\ref{fig:emission_mech} we provide a practical, heuristic guide of differentiating between the different radio emission mechanisms operating in stellar systems. 

In contrast, radio emission from the magnetised Solar System planets is dominated by auroral processes. What this loosely refers to is emission that is generated close to the magnetic poles of the planet. A key ingredient to driving auroral emission is electron acceleration. The injection of a high velocity electron population into a quiescent magnetosphere powers bright, coherent radio emission via the electron cyclotron maser (ECM) instability \cite{Wu1979,Treumann2006}. ECM emission is highly circularly polarised and occurs at the local cyclotron frequency. The maximum frequency of ECM emission is directly proportional to the ambient magnetic field strength at the emitting point. Therefore, ECM emission is a direct probe of the magnetic field strength of the emitting body \cite{1985ARA&A..23..169D}. Another characteristic of ECM emission is that it is beamed, in that the radio waves propagates outwards on the surface of a cone that is near-perpendicular to the local magnetic field. A result of this beaming is that the emission is only generally visible for brief windows in time and can exhibit complicated periodicity \cite{Kavanagh2023}.

Several distinct mechanisms can produce the population of accelerated electrons in planetary magnetospheres. The first is due to the interaction of the solar wind directly with a planet's magnetosphere. In this case, the incident solar wind is magnetised and compresses the planet's intrinsic magnetosphere, resulting in magnetic reconnection on the nightside of the planet. This process is often referred to as the Dungey cycle \cite{dungey61}, and interaction of the high energy electrons with the atmosphere produces the aurora australis and borealis on Earth. Interestingly, the observed auroral radio power from the magnetised Solar System planets is seen to directly scale with the power of the incident solar wind, both  kinetically and magnetically  \cite{Zarka2007,Desch1984,Farrell1999,Zarka2001} -- with the kinetic relation often referred to as the radiometric Bode's law, and the magnetic relation as the radio-magnetic scaling law.

Despite being the brightest low-frequency radio emitter in the Solar System, Jupiter's auroral emission deviates from this narrative as two different processes dominate the auroral radio emission: the breakdown of co-rotation and the Jupiter-satellite interaction \cite{Nichols2011,Saur2004}. For the breakdown of co-rotation model, Jupiter's inner magnetosphere is continuously supplied with plasma from Io's volcanic outgassing. The inner plasma is forced to co-rotate with the magnetic field of Jupiter as it is centrifugally expelled outward. At a certain distance in Jupiter's magnetosphere, the magnetic field is not strong enough to force co-rotation of the plasma, producing a shear against the outer plasma environment. This shearing generates a current system in which electrons are accelerated from the equator to the poles, producing the auroral ring \cite{Terasawa1978,Zarka1983,Genova1987}. 

The brighter and more localised radio emission from Jupiter is associated with the magnetic field lines of Jupiter connecting it to its Galilean moons, particularly to Io \cite{bigg64, Neubauer1980, Saur2004, Marques2017}. The driving mechanism of this emission is thought to either be due to Alfv\'en waves\footnote{Alfv\'en waves are transverse plasma waves that travel along magnetic field lines. They are produced when ions oscillate in response to the restoring force provided by the tension of magnetic field lines.} generated by the perturbation of Jupiter's magnetic field by the inner moons \cite{Neubauer1980}, or magnetic reconnection occurring between the magnetic fields of Jupiter and the moons \cite{ip04, lanza12,Turnpenney2018}. In either case, electrons are accelerated, subsequently producing bright circularly-polarised radio emission via the ECM mechanism. A constraint on this form of emission is that it can only be produced if the perturbing body orbits inside the Alfv\'en surface\footnote{The Alfv\'en surface is defined as the three-dimensional boundary that separates a star's corona from its stellar wind -- the boundary at which information in the stellar wind can not propagate back to the surface of the star. This is the locus where the Alfv\'{e}nic Mach number is unity.} of the host body for star-planet interactions, a region where the host body's magnetic field dominates (see Section\,\ref{sec:spi} for further details). 

The observed radio emission processes from Jupiter serve as archetypes for what is expected from extrasolar planetary systems, where it is possible to produce emission that is orders of magnitude brighter than that observed in Solar System by scaling the interaction by mass, obstacle size, magnetic field strength of the stellar wind and of the planetary obstacle, and orbital distance (see Section\,\ref{sec:spi} for quantitative  expressions of this statement) \cite{Zarka2007,Zarka2001,Saur2013}. These interactions are often referred to as magnetic star-planet interactions in the literature, which will refer to generally as star-planet interactions (SPI) in this manuscript. 

\begin{figure}
    \centering
    \includegraphics[width = \textwidth]{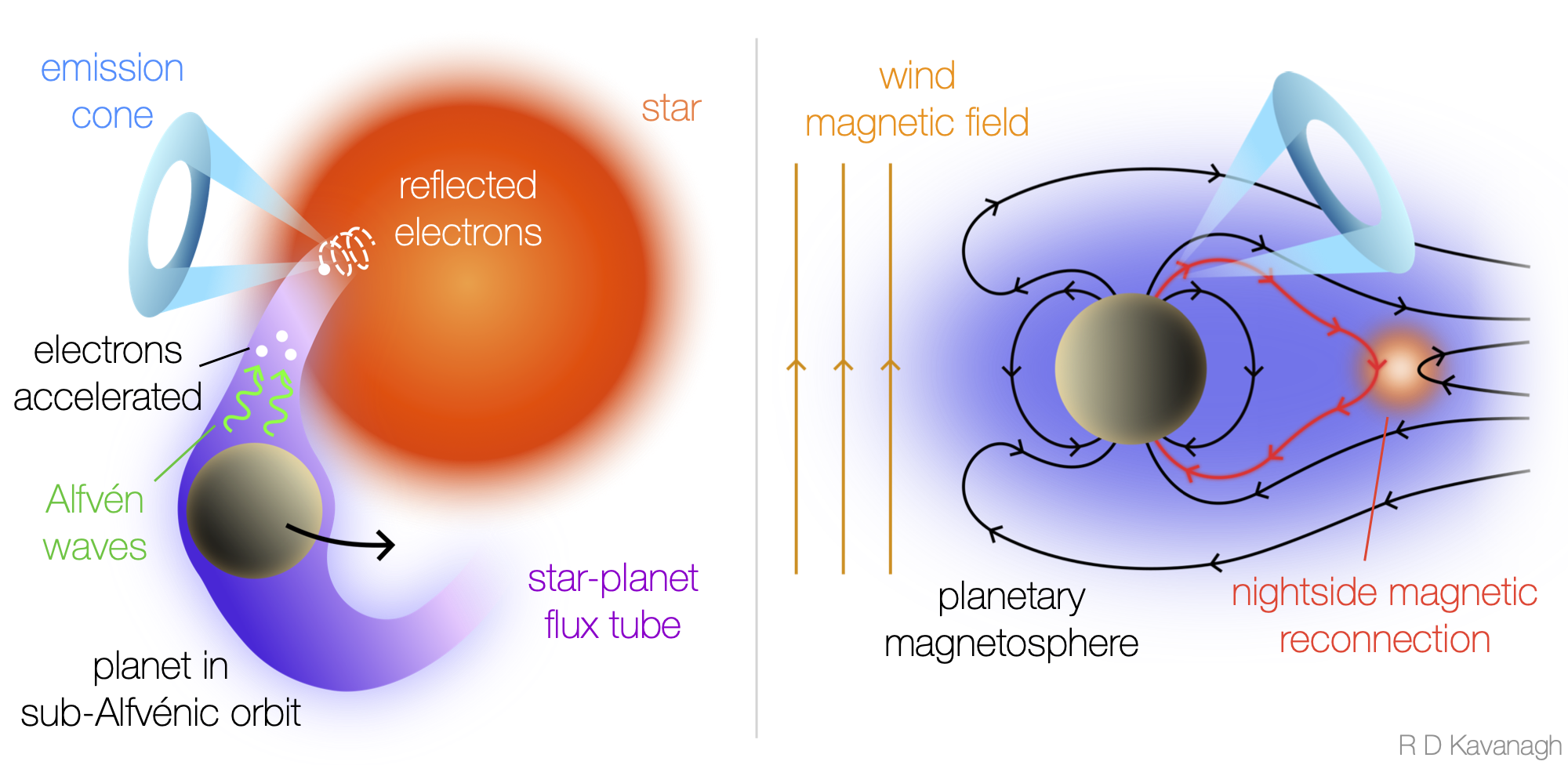}
    \caption{Sketch illustrating the two putative sources of electron cyclotron maser (ECM) emission in exoplanetary systems. \textit{Left Panel:} Emission induced on a star by a close-in planet. If the planet orbits inside the Alfv\'en surface of the star, it can perturb the star's magnetic field, producing Alfv\'en waves which propagate back towards the star. These waves interact with electrons, accelerating them towards the stellar surface. Electrons with sufficiently large pitch angles undergo a magnetic mirroring effect and are reflected, producing ECM emission which propagates in a hollow cone. \textit{Right Panel:} Auroral emission induced on a planet by the interaction of its magnetosphere with the incident wind of its host star. The magnetic field carried by the stellar wind causes the planet's magnetosphere to open up on the dayside (left). These field lines are pushed towards the nightside (right), where they subsequently reconnect. The energy released in the magnetic reconnection accelerates electrons back towards the poles along the field line highlighted in red, where they reflect and power ECM emission in a similar manner to that described for the left panel. For clarity, the emission cone is only shown for the Northern hemisphere in both cases.}
    \label{fig:sketch}
\end{figure}

As illustrated in Figure~\ref{fig:sketch}, there are two sub-types of SPI that are relevant in producing ECM radio emission from radio stellar systems that contain a low-mass host star:

\begin{enumerate}

\item \textbf{Sub-Alfv\'{e}nic interactions:} \\
Analogous to the Jupiter-Io interaction. Radio emission occurs along the stellar magnetic field line connecting the star and planet, driven by interactions between the field line and the planet. It requires the planet to be orbiting inside the Alfv\'en surface, with either an intrinsic or induced magnetic field. It is sometimes referred to as magnetic star-planet interactions. The power produced by this interaction appears to only follow the radio-magnetic scaling law \cite{Zarka2018}.\\

\item \textbf{Wind-magnetosphere interactions: }\\
Auroral radio emission is produced in a ring-like configuration near the planet's magnetic poles, driven by the interaction of the incident stellar wind and planetary magnetosphere. The power produced by this interaction appears to follow both the radiometric Bode's law and the radio-magnetic scaling law for most auroral sources in the Solar System.

\end{enumerate}

Naturally, the SPI terminology is no longer accurate once either the star or planet are replaced with a brown dwarf or moon. However, the underlying physical processes are still relevant in those scenarios. For instance, an exomoon could produce detectable radio emission on an exoplanet via a sub-Alfv\'enic interaction. Stochastic flaring and breakdown of co-rotation \cite{nichols12} are also pathways to generate coherent radio emission in these systems, however since these processes do not explicitly rely on the presence of a planet, they do not fall under the umbrella of SPI. We also note that the term SPI often more broadly also refers to the irradiation of the planetary atmospheres by the host star's light and its subsequent interactions with the wind of the host star \cite{lecavelier12} and tidal interactions between the two bodies \cite{cuntz00}. However, neither of these processes are expected to directly produce bright signatures in the radio, therefore we do not discuss them in this manuscript. 

\section{Stellar Flares and Coronal Mass Ejections}

Radio observations of the Sun and other solar-like and low-mass stars have shown that stellar flare production involves all layers of the stellar atmosphere, from the photosphere to the corona. The large extent of physical scales involved in producing flares implies that a range of processes are responsible, such as coronal heating and particle acceleration \cite{BenzGudel2010}. A consensus picture is emerging that connects the radio, optical, ultraviolet, and X-ray flare luminosities of stars to coronal heating. However, one remaining challenge is using time-resolved, simultaneous multiwavelength observations\cite{2022ApJ...938..103H} to trace coronal mass ejections, both to understand stellar mass loss and to directly probe the space weather environments faced by exoplanets. 

There is a well-established model for the non-thermal incoherent radio emission from solar and stellar flares. The model is motivated by a remarkable correlation between the quasi-steady gigahertz-frequency radio $L_{\nu,\mathrm{rad}}$ and soft X-rays $L_{\mathrm{X}}$ luminosities: $L_{\mathrm{X}} \propto L_{\nu,\mathrm{rad}}^{0.73}$ \cite{Guedel1993ApJ...405L..63G, Benz1994A&A...285..621B}. This correlation is often referred to as the G\"udel-Benz relation. Such a relation is canonically explained via the fact magnetic reconnection events produce streams of non-thermal, accelerated coronal electrons. These streams impact the chromosphere and cause evaporation of million-degree plasma, depositing their kinetic energy as thermal energy in the corona \cite{Antonucci1984ApJ...287..917A}. The free electrons produced by the magnetic reconnection event are observationally traced by the gigahertz-frequency gyrosynchrotron emission as they radiate in the stellar magnetic field, while the heated plasma is identified by its thermal soft (0.2-2\,keV) X-ray emission. It is important to note that while the explanation presented here is qualitatively plausible, it has difficulties in quantitatively describing the G\"udel-Benz relation \cite{Airapetian1998}. The G\"udel-Benz relation equally applies over many orders of magnitude to most solar and stellar flares, suggesting that very hot, active stellar coronae are heated by the integrated energy of flares -- including those flares that may be too low energy to detect individually \cite{2020IJAsB..19..136A,2022ApJ...924..115O}. 

Small scale, stochastic flares could indeed be responsible for the quasi-steady soft X-ray and low circularly polarised, steady gigahertz-frequency radio emission from stellar systems. If the energy distribution $E$ of stellar radio flares follows a power law, low energy flares will be largely responsible for heating the corona if the power-law index $\alpha$ is $>2$ in $dN/dE \propto E^{-\alpha}$. However, this has been challenging to test in the radio: observational flare distributions are limited at low $E$ by the instrumental detection limit, depends on stellar distance and flare peak luminosity, and long time series are required.

At short UV, optical, and the non-thermal hard X-ray ranges, Solar and stellar flare statistics do indeed reveal power-law distributions down to the light curve noise, mostly with $\alpha \gtrsim 2$ for active stars  \cite{Crosby1993SoPh..143..275C, Audard1999ApJ...513L..53A, Audard2000ApJ...541..396A, Kashyap2002ApJ...580.1118K,  Guedel2003ApJ...582..423G, Arzner2004ApJ...602..363A, Stelzer2007A&A...468..463S,Maehara2012Natur.485..478M, Aschwanden2015ApJ...814...19A, Yang2019ApJS..241...29Y}. These optical stellar flare statistical studies have been revolutionised since the \textit{Kepler} \cite{Borucki2010} and TESS \cite{Ricker2015} space telescopes have produced hundreds of thousands of high-cadence, high-precision broadband optical light curves. Homogeneous statistical catalogs of flares \cite{Walkowicz2011,Hawley2014,Davenport2016,Yang2019,Gunther2020,Feinstein2020,Gao2022,Pietras2022} have been used to investigate the population level statistics \cite{Davenport2019,Feinstein2022b}, flares characteristics of planet-hosts \cite{Howard2022,Gilbert2022,Feinstein2022}, and have revealed Carrington-analog superflares on solar-like stars \cite{Shibayama2013,Shibata2013,Maehara2015,Notsu2013,Notsu2019,Cliver2022}. 

New radio facilities are for the first time capable of probing whether large optical flares are accompanied by coronal mass ejections. While this is true for the largest of the Sun's flares \cite{Yashiro2006}, the commonly flaring M~dwarfs have different magnetic geometries and stronger global fields \cite{Morin2010,DonatiLandstreet2009}. Recent modelling efforts \cite{AlvaradoGomez2019} have shown that large overlying magnetic fields can prevent mass breakout, resulting in the occurrence of confined flares,
which do not have an accompanying eruption. The ongoing, all-sky TESS mission uniquely complements modern radio telescopes for multiwavelength flare studies: simultaneous ASKAP and TESS observations have traced a radio burst from Proxima~Centauri \cite{Zic2020}, and also radio emission with no optical counterpart \cite{Rigney2022}. The TESS flare rates of LOFAR-detected stars are correlated with their X-ray luminosities, consistent with the G\"{u}del-Benz mechanism, but several show high ECM radio luminosity and few or no flares or X-ray emission. Such a deviation has been taken as evidence in support of the radio emission being generated by star-planet magnetic interaction since radio emission generated by plasma or gyrosynchrotron mechanisms are expected to correlate with magnetic events that often have an optical/X-ray flare counterpart \cite{Pope2021}.

While Type~II bursts are the only unambiguous radio proof of CME material escaping from the stellar magnetosphere, there are no firm detections of Type~II bursts from other stars despite significant observational effort \cite{Crosley2016ApJ...830...24C, Crosley2018ApJ...862..113C, Villadsen2019ApJ...871..214V,Callingham_crdra}. The non-detections of extrasolar Type~II bursts may be due to sensitivity limitations, magnetic confinement of CMEs on radio-bright stars \cite{Alvarado-Gomez2018ApJ...862...93A}, or an Alfv\'en speed that prevents shock formation \cite{Alvarado-Gomez2020ApJ...895...47A}. Instead, radio observations of active M~dwarfs have found other types of coherent radio bursts, including hours-long events attributed to ECM emission \cite{Villadsen2019ApJ...871..214V,Zic2019MNRAS.488..559Z,Callingham_crdra,Bastian22} that do not have a direct Solar analogue. Other promising evidence of stellar CMEs \cite{Osten2022IAUS} includes blueshifts of chromospheric lines and EUV/X-ray coronal dimming \cite{veronig21}, but such measurements do not inform us if the material was retained in the stellar magnetosphere -- implying the plasma never impacts a putative planet. Low-frequency, wide-field surveys with $<10$\,mJy sensitivity on timescales of minutes are the most likely avenue to detect an extrasolar Type~II burst and confirm that mass has been ejected since such bursts are stochastic and the emission frequency is likely lower in weaker magnetic field strengths \cite{1985ARA&A..23..169D,guedel2002}. Such a detection would allow a measurement of the plasma density as the radio emission occurs at the plasma frequency $\nu_{p} \propto n_{e}^{1/2}$, where $n_{e}$ is the electron density, allowing us to trace the particle flux at the point of impact of an exoplanet \cite{vidotto18}.

Finally, the Sun exhibits steady mass loss through the solar wind. Measuring stellar winds have been notoriously difficult. Radio observations of thermal bremsstrahlung emission have so far only placed upper limits on stellar winds from young solar analogs \cite{fichtinger17,ofionnagain19}, and indirect ultraviolet wind measurements of M~dwarfs suggest a fairly large spread of inferred mass loss rates with surface X-ray flux \cite{Wood2021}. There is promise with the Next Generation Very Large Array (ngVLA; \cite{DiFrancesco2019}) to be able to detect mass loss from the steady stellar wind of nearby M~dwarfs \cite{OstenCrosley2017}. 

\section{Radio Emission Directly from Exoplanets}

Exoplanetary systems are observed in the radio primarily to search for auroral emission powered by breakdown of co-rotation, wind-magnetospheric interactions, or emission induced on the host star by an exoplanet via sub-Alfv\'enic interactions. However, the energetics of these different models imply any radio detection of an exoplanet will likely be near the sensitivity limits of current radio facilities \cite{Zarka2007,Nichols2011,Lynch2017,Griessmeier17PREVIII}. Therefore, the direct detection of exoplanetary auroral radio emission has been elusive despite decades of searching \cite{Zarka2015SKA,G2015,Griessmeier17PREVIII}. Following several foundational works \cite{,Zarka2007,Farrell1999,Zarka2001,Zarka1997pre4}, a large collection of theoretical work has been published (e.g. \cite{Lazio2004,Stevens2005,Jardine2008,Hess2011,Nichols2011,Vidotto2012,See2015,Vidotto2015,Nichols2016,Weber2018,Lynch2018,Wang2019,Kavanagh2019,Turnpenney2020}) predicting the intensity and frequency of the radio emission of exoplanets observed from Earth. 

As stated in Section\,\ref{sec:sun_jup}, radio emission directly from an exoplanet can be powered by the energy deposited on the planetary magnetosphere from the incident stellar wind \cite{Farrell1999}, or via a sub-Alfv\'enic interaction with a moon \cite{Noyola2016,Narang2023,Narang2023b}. The radiometric Bode's law \cite{Farrell1999,Lazio2004} and the radio-magnetic scaling law \cite{Zarka2007,Zarka2001,Gr2007} have been used to predict the radio intensity of exoplanets, directly relating the incident stellar wind power (kinetic or magnetic) to the emitted radio power. However, the uncertainty on the predicted flux density can be greater than an order of magnitude, and the uncertainty on the predicted maximum emission frequency can be off by a factor of 2-3 \cite{Gr2007}. Therefore, these predictions should be used with caution. 

A large number of observational campaigns have been performed to search for radio emission from exoplanets \cite{lazio2024,Zarka2018hoe}, resulting in clear non-detections  \cite{Yantis1977,Winglee1986,Bastian2000,Lazio2007,Smith2009,Lazio2010a,Hallinan2013,2015MNRAS.446.2560M,Lynch2017,Lenc2018,Lynch2018,2018A&A...612A..52O,Gasperin2020,2022AJ....163...15C,2023ApJ...952..118R,bloot2024} and a few tentative but contested detections \cite{Lecavelier2013,Sirothia2014,Vasylieva2015,Turner2021,Pineda2023}. No claim of detection has yet been confirmed by follow-up observations. The causes for radio non-detections directly from exoplanets are degenerate \cite{Zarka2015SKA,G2015,Griessmeier17PREVIII,kavanagh19}. Namely, it is possible that we have no clear detections because: 

\begin{enumerate}

\item Observations have not been sufficiently sensitive;
\item The planetary magnetic field is too weak, implying ECM emission is not produced at accessible observing frequencies;
\item Earth was outside the beaming pattern of the radio emission at the time of the observation \cite{Hess2011,Ashtari2022};
\item The ECM conditions are varying in response to variations in the stellar wind conditions and/or the electron velocity distribution, causing the radio emission to drop below the detection threshold of the observation \cite{Fischer2019,Elekes2023};
\item Inhibition of the ECM process from, for example, inflated ionospheres (e.g. \cite{Weber2017, Weber2017pre8, daley-yates18, Weber2018, Erkaev2022}); 
\item The wind of the host star absorbs or prevents the escape of the emission\cite{kavanagh19}, 
\item The presence of high plasma densities in the planetary magnetosphere or surrounding stellar wind suppresses wave propagation \cite{1985ARA&A..23..169D,Weber2017pre8}, or;
\item Difficulties disentangling whether the detected coherent emission is stellar or exoplanetary in origin since stars also can intrinsically produce coherent emission (e.g. \cite{Callingham_crdra,bloot2024}).

\end{enumerate}

\noindent It is possible all of these effects are in play for each individual system, though their relative contributions have not yet been quantified -- implying it is difficult to infer physically meaningful information about a single system from a non-detection of ECM emission.

Progress in detecting radio emission directly from an exoplanet will largely come from complete phase coverage of the orbit of the exoplanet, preferably several times, and increased sensitivity at low frequencies. Further development of sensitive instruments that observe at $\lesssim300$\,MHz is vital if we want to probe systems at Jupiter-mass or below since ECM emission scales linearly with the magnetic field strength of the body, and from dynamo theory it is unlikely planets with mass less than 10 times that of Jupiter will possess magnetic field strengths significantly exceeding $\sim$100\,G \cite{brian2024}. For example, Jovian radio emission cuts off above $\approx$40\,MHz since the its magnetic field does not exceed $\approx$14\,G \cite{Connerney2022}).

\section{Auroral Radio Emission from Ultracool Dwarfs}
\label{sec:ucds}

Due to the difficulty in conclusively detecting radio emission directly from exoplanets, we can leverage the considerable effort that has gone into observing ultracool dwarfs (UCDs; spectral types $>$M7), to examine magnetic processes across the substellar regime, which is expected to have the same underlying physics as direct emission from exoplanets and SPIs. Some UCDs have exhibited bright radio bursts \cite{Berger2001Natur.410..338B} generated by ECM emission \cite{Hallinan2007ApJ...663L..25H}, allowing us to trace auroral processes in magnetospheres more similar to Jupiter than the Sun \cite{Hallinan2015Natur.523..568H, Kao2016ApJ...818...24K, Pineda2017ApJ...846...75P}. The Jupiter-like analogy for UCDs has recently gained further support with the detection of synchrotron radiation belts around a UCD \cite{Kao2023,2023arXiv230306453C}, a key radio morphological characteristic possessed by Jupiter's magnetosphere. 

The first serendipitous discovery of bursting radio emission from a brown dwarf \cite{Berger2001Natur.410..338B} and ensuing UCD radio detections, demonstrated that their radio activity can strongly depart from well-established stellar coronal/flaring relationships \cite{Berger2005ApJ...627..960B, Williams2014ApJ...785....9W}. In light of the subsequent discovery that UCD radio bursts can exhibit periodic timing \cite{Hallinan2006ApJ...653..690H}, their strong circular polarization and high brightness temperatures implies the ECM process is at least partly responsible for the radio emission from some UCDs \cite{Osten2005ApJ...621..398O,Hallinan2007ApJ...663L..25H}. Despite thorough gigahertz-frequency radio searches of UCDs, detections rates have remained persistently low $\sim$10\% detection rates \cite{Osten2006ApJ...644L..67O,Berger2010ApJ...709..332B,Antonova2013A&A...549A.131A, Burgasser2013ApJ...762L...3B, Route2012ApJ...747L..22R, Route2013ApJ...773...18R, Route2016ApJ...830...85R, Lynch2016MNRAS.457.1224L,Kao2023b}.

Periodically bursting radio emission traces extrasolar aurorae in UCDs. Simultaneous radio and optical observations of the M8.5 dwarf LSR~J1835+3259 showed that the impacting electron current traced by its bursting radio emission creates a surface feature that is spectrally distinct from typical magnetic spots seen on stars \cite{Hallinan2015Natur.523..568H}. UCDs with optical-infrared periodic variability or H$\alpha$ emission can have an 80\% radio detection rate \cite{Kao2016ApJ...818...24K}. Magnetospheric current systems modeled off of Jupiter can explain the generally correlated radio and optical/atmospheric behaviors across the UCD population \cite{Pineda2017ApJ...846...75P,Vedantham2020ApJ...903L..33V}. 
In this picture, brown dwarfs are magnetic analogues to gas giants and laboratories for studying planetary dynamos \cite{Kao2016ApJ...818...24K, Kao2018ApJS..237...25K,Vedantham2023}. Furthermore, the discovery of hitherto unknown brown dwarfs via their radio emission \cite{Vedantham2020ApJ...903L..33V,Rose2023,Vedantham2023} demonstrates that large-field radio surveys can access a new discovery space for brown dwarfs or planets that other observational means may overlook. In particular, it is expected the next generation of wide-field radio surveys will likely identify the lowest mass or most distant UCDs, since infrared surveys are reasonably complete within 25\,pc for UCDs of spectral types earlier than T4 \cite{best2024}.

The broad success of the UCD auroral paradigm naturally leads to a substantive open question: what is the source of the magnetospheric plasma in such systems? This magnetospheric condition may determine the overall low radio detection rate ($<10\%$) \cite{Lynch2016MNRAS.457.1224L,Route2016ApJ...830...85R}. A potentially viable solution is exoplanet companions \cite{Pineda2017ApJ...846...75P, Kao2019MNRAS.487.1994K}. In the same way that Io seeds the Jovian magnetosphere with plasma, these satellites could provide the plasma for their UCD hosts. Although challenging, searches for planets around UCDs are ongoing \cite{Delrez2018SPIE10700E..1ID,Tamburo2022AJ....163..253T, Limbach2021ApJ...918L..25L}, with TRAPPIST-1 \cite{Gillon2017Natur.542..456G} providing a prototypical example. The radio emission itself is also enabling companion searches, and astrometric radio monitoring via very long baseline interferometry has already yielded evidence for a companion around an auroral UCD \cite{Curiel2020AJ....160...97C}.

A statistically significant coincidence between UCD planetary systems and bright radio emission remains unproven. Nevertheless, because a satellite population likely exists, the strong magnetic fields of the lowest-mass stars and brown dwarfs \cite{Kao2018ApJS..237...25K} make them exceptional candidates for searching for magnetic host-planet interactions \cite{Saur2021}. Alternatively, the radio emission could be similar to the breakdown of co-rotation seen in the Jupiter system, with the plasma supplied by magnetic reconnection events in the UCD's magnetosphere. Long term radio monitoring of the systems, and determination of the UCDs rotation period, is required to determine which model is correct since the periodicity will either correlate with the rotation of the UCD or the orbit of a putative satellite.

\section{Radio Emission from Star-Planet Interactions}\label{sec:SPI}

As well as radio emission directly from a planetary magnetosphere, it is thought to be possible to drive radio emission via the magnetic connection between a close-in planet and its star  -- \`{a} la a scaled-up Jupiter-Io interaction \cite{Zarka2007,Zarka2001,Gr2007,Zarka2018haex,Griessmeier2018haex}, with a star taking the place of Jupiter and a planet taking the place of Io.

Radio emission from SPI via ECM emission can potentially reach higher frequencies than planetary magnetospheric radio emission (up to gigahertz frequencies for some M~dwarfs, massive, or young stars due to their relatively strong magnetic fields) since the auroral emission is occurring in the magnetosphere of the star. This higher frequency of emission makes radio from SPI easier to detect with ground-based radio observatories than that expected directly from exoplanetary magnetospheres. Note that the radio emission from SPI only gives a direct measure of the magnetic field of the star, not the planet. However, potentially it is possible to model a planet's magnetic field in such an interaction (see, e.g. \cite{G2015,Preusse06,Kopp2011}), similar as is done for Ganymede \cite{2020GeoRL..4790021L}. Emission from SPI may also manifest as variation in X-rays \cite{Scharf2010} and ultraviolet/optical/near-infrared activity indicators \cite{Shkolnik2005,Shkolnik2008,Cauley2019,Klein2022} -- but all claims of detection of SPI at these wavelengths are disputed \cite{2011ApJ...735...59P,2013A&A...552A...7S,2015ApJ...799..163M,2020AJ....159..194V}.

Similar to radio emission from exoplanets, there has been no confirmed detection of radio emission from SPI. However, some recent publications show encouraging signals that merit follow-up. LOFAR observations have shown low-frequency radio emission from the quiescent M dwarf GJ\,1151, and several others, at 144 MHz \cite{Vedantham2020,callingham21}, which has been attributed to the interaction of the star with a close-in Earth-size planet due to the emission and stellar properties. An intensive campaign of precise radial velocity observations with HARPS-N \cite{Pope2020}, HPF \cite{Mahadevan2021} and CARMENES \cite{Perger2021} has detected a $> 10.6\,M_\oplus$ companion to GJ\,1151 in a 390\,d orbit at a separation of 0.57\,au, but only a 1.2$M_\oplus$ upper limit on the mass and orbit of any companion close enough to be in the sub-Alfv\'{e}nic region of the star \cite{Blanco-Pozo2023}. 

Several other recent unconfirmed SPI detections also merit followup with intensive observing campaigns, particularly in synergy with spectropolarimetric monitoring aimed at modelling the large-scale field of the host stars. At gigahertz frequencies there have been suggested detections of ECM emission possibly modulated by planetary orbits from Proxima Centauri \cite{PerezTorres2021} and YZ~Ceti \cite{Pineda2023,Trigilio2023}. Modulation of optical activity tracers close to the period of AU\,Microscopii\,b has been interpreted as SPI \cite{Klein2022}, and suggests AU\,Mic as a promising target for SPI. However, only evidence of stellar rotation modulation in the radio has been detected from AU\,Mic \cite{bloot2024}.

\subsection{Theory of magnetic star-planet interactions}\label{sec:spi}

Considering the likelihood of detection of radio emission from SPI with the current generation of instruments, in this section we outline the theory of SPI and how it can be used to derive physical characteristics of SPI systems.

SPI radio emission is produced by the ECM process near the local electron cyclotron frequency $\nu_{c}$, which peaks at $\nu_{c} = 2.8 B_{*}$\,MHz, where $B_{*}$ is the magnetic field strength of the star in Gauss. Such emission is produced in a rarefied ($\nu_{c} >> \nu_{p}$) plasma by an unstable electron population with characteristic energy of $\sim$1-20\,keV \cite{Treumann2006,Zarka1998,Lamy2018,Louis2019,Sulaiman2022,Lamy2022}.

Furthermore, SPI radio emission is predominately emitted via the extraordinary (x) magneto-ionic mode, implying the polarisation of the emission is highly circular or weakly elliptical. The emission is beamed at a large angle from the magnetic field, typically $60^\circ-90^\circ$, and in a thin conical sheet of $1^\circ-2^\circ$ thickness \cite{Zarka1998}. We schematically show such emission geometry in Figure~\ref{fig:sketch}. Modelling the ECM emission often involves assumptions about the stellar magnetic field topology, making it possible to produce a `visibility curve' for radio emission for a SPI system \cite{kavanagh19,Fares2010}. In the limit of a dipolar stellar magnetic field, radio emission from SPI can have a very low duty cycle ($<10\%$) and a sensitive dependence on the obliquity between the axes of the stellar dipole, rotation, and planetary orbit \cite{Kavanagh2023}.

The radio power produced by variable ECM emission is hard to predict from first principles: ECM growth rates and the wave path along which it operates depend on the details of the distribution of the unstable $\sim$1-20 keV electrons in the auroral regions, and on the magnetic field topology and ambient plasma distribution in the radio sources. However, observations collected in the Solar System reveal that the emitted auroral radio power $P_r$ averaged over time, frequency, and solid angle is approximately proportional to both the incident kinetic $P_\mathrm{kin}$ and Poynting $P_B$ fluxes \cite{Zarka2007,Zarka2018hoe}:

\begin{equation}
P_r = \alpha P_\mathrm{kin} = \alpha \rho v^3 \pi {R_\mathrm{obs}}^2 
\end{equation}

\begin{eqnarray}
    P_r = \beta P_B = \beta \frac{B_\perp^2}{\mu} v \pi {R_\mathrm{obs}}^2
    \label{eq:P_B},
\end{eqnarray}

\noindent
where $\rho$ is the density of the wind incident on the planet's magnetosphere, $R_\mathrm{obs}$ is the effective size of the obstacle, $v$ the incident flow velocity in the obstacle's frame, $B_\perp$ the magnetic field in the flow perpendicular to $v$, and $\mu$ is the permeability of space. The proportionality factors $\alpha$ and $\beta$ are the overall efficiencies of the conversion of the kinetic and Poynting flux into the emitted radio power, which are empirically estimated to be of order of $10^{-5}$ and $10^{-3}$ respectively from Solar System observations \cite{Zarka2001,Zarka2018}. 
When auroral and satellite-Jupiter radio emissions are considered together, only equation Equation~\ref{eq:P_B} holds. Its high efficiency ($\beta \sim 10^{-3}$) explains why the Jupiter-Io interaction is the brightest object in the sky below 40\,MHz, outshining even the quiet Sun.

The detailed magnetohydrodynamics of SPI models lead to a wide range of estimates of the Poynting flux. For sub-Alfv\'{e}nic interactions we can have 1) force-free models, where the electric current and magnetic field vectors are parallel \cite{Lanza2009} or; 2) Alfv\'en wing models \cite{Saur2013,Strugarek2021} with pairs of standing waves in the magnetised wind, which separately may connect the planet either to the star or interplanetary space.

For the Alfv\'en wing model, the power transmitted in one Alfv\'en wing was first estimated as $P_{\rm AW} = \epsilon P_B$, where $\epsilon=(1+M_A^{-2})^{-1/2}$. $M_A$ is the Alfv\'en Mach number, which is defined as the plasma flow speed divided by the Alfv\'en velocity. For small $M_A$, $\epsilon \approx M_A$ \cite{Zarka2007,Zarka2001}. For reference, $M_A = 0.3\pm0.1$ for Io's interaction \cite{Kivelson2004}. Based on a nonlinear solution of Alfv\'en wings \cite{Neubauer1980}, the total power can be expressed in closed form for small $M_A$ as \cite{Saur2013}
\begin{eqnarray}
  P_{AW} = 2  \pi R^2 \bar{\alpha}^2 M_A  \frac{B_\perp^2}{\mu_0} v_0 \;.
  \label{e:P}
\end{eqnarray}

\noindent
Here $\bar{\alpha}$ is the interaction strength  ($0 \le \bar{\alpha} \le 1$) and $R$ is the radius of the central Alfv\'en tube carrying the energy away along the background field. It is often approximated by the effective radius of the obstacle \cite{Saur2013}. $\bar{\alpha}$ is expected to be close to one if the planet possesses a dense atmosphere/ionosphere. This Alfv\'{e}n wing expression is widely used \cite{Turnpenney2018,Fischer2019,Strugarek2015}, and is a more exact solution than the early estimate $\approx M_A P_B$ presented by \cite{Zarka2007}. The energy flux in Equation~\ref{e:P} is a factor of two larger as than that presented by \cite{Zarka2007} as it considers the energy fluxes inside and outside the central Alfv\'en tube defined by $\pi R_{obs}^2$ \cite{Saur2013}. The dependence on $M_A$ also highlights that low density stellar winds will have a lower amount of power emitted compared to high density stellar winds. 

In force-free models of SPI powered by reconnection, the energy fluxes have been estimated as $P_{FF} = \gamma P_B$ with the efficiency factor $0 \le \gamma \le 1 $ \cite{Lanza2009}. These fluxes are roughly the same order as those in the Alfv\'en wing models. In force-free models of the Jupiter-Io interaction, such as the unipolar inductor, the energy fluxes are controlled by the conductance of Jupiter's ionosphere and not by the wave impedance of the Alfv\'en waves \cite{Goldreich1969,Saur2021}.
 
One commonly used form of a force-free model is called the stretch-and-break model \cite{Lanza2013,Strugarek2022}, where the planetary polar magnetic field $B_P$ is stretched and energy is released by reconnection. In this case the power produced is

\begin{align}
       P_{SB} &= 2  \pi R_{P}^2  \frac{B_P^2}{\mu_0} v_0 f_{AP},  
    \label{eq:P_SB}
\end{align}

\noindent where $R_P$ is the radius of the planet and $f_{AP}$ is the area fraction of the planet where magnetic field lines connect to the stellar wind. $f_{AP}$ might have a typical value of order 0.1 depending on the strength and configuration of the planetary and wind magnetic fields. The direct occurrence of the planet's internal magnetic field $B_P$ is the primary difference between the Alfv\'{e}n and stretch and break model. It would mean that $P_{SB}/P_{AW}$ can be 100-1000 \cite{Strugarek2022}. However, the Poynting flux in Equation\,\ref{eq:P_SB} is primarily perpendicular and not directed along the flux tube towards the star, and the flow velocities $v_0$ over the poles will be reduced as well. The fluxes in Equation\,\ref{eq:P_SB} have neither been seen in numerical simulations \cite{Strugarek2015}, nor take place at Jupiter's magnetized moon Ganymede. We suggest further work on the applicability of  Equation\,\ref{eq:P_SB} is necessary if the stretch and break model is to be used for predictions.

A general prediction from all of these models is that the brightest radio emission from SPI will be produced by the largest planet as close as possible to a star with the strongest magnetic field. This heuristic can inform future SPI searches but needs to be used with caution as such a heuristic neglects potentially important second order effects, such as ECM inhibition due to inflated ionospheres \cite{Weber2018,Weber2017, Weber2017pre8, daley-yates18, Erkaev2022}. 

\subsection{Stellar wind environments around planet-hosting stars}
\label{sec:spi4}

Whether or not radio emission from SPI occurs, as well its morphology and strength, depends on the plasma environment that the planet is embedded. We require detailed knowledge of this plasma if we are to predict and interpret signatures of SPI. For low-mass stars, the primary source of the interplanetary plasma is its stellar wind, which expands outward from the stellar surface, likely via a combination of magnetic and thermal pressure forces \cite{shoda20}. Additional sources may further add to the interplanetary environment, such as atmospheric mass loss from the planet itself \cite{fossati13} and CMEs \cite{osten15}.

Unlike our Sun, we lack detailed knowledge of the stellar wind environments around other low-mass stars \cite{vidotto21}. This is primarily because their winds are very tenuous, implying they do not produce strong observational signatures. However, sophisticated magnetohydrodynamic (MHD) models can be deployed to obtain three-dimensional snapshots of the winds of low-mass stars. Measurements of the mass-loss rates of the winds of low-mass stars are currently limited to a handful of cases, and rely on indirect methods (see, \cite{fichtinger17, wood04, jardine19}). Nevertheless, this has not stopped trends relating mass-loss rates to more readily available information such as surface X-ray fluxes being established \cite{vidotto18}. In turn, these trends are used to inform MHD models.

Another key constraint that can be implemented into MHD models of the winds of low-mass stars are reconstructed surface magnetic field maps. Since the inception of the ZDI method \cite{semel89}, such data have become readily available for a relatively large population of low-mass stars \cite{Morin2010,Klein2022}. The magnetic field is embedded in the stellar wind and strongly influence the dynamics of the flow \cite{kavanagh19}. Additionally, the wind magnetic field is essential in predicting and interpreting signatures of star-planet interaction for at least two reasons. Firstly, it is necessary to estimate the location of the Alfv\'en surface, and thus predict whether the planet has a sub- or super-Alfv\'{e}nic orbit \cite{kavanagh21}. Secondly, the local magnetic field is also required to estimate the Poynting flux reaching the planet, which dictates the strength of the interaction \cite{fischer19}. By implementing such maps as boundary conditions in MHD simulations of stellar winds, we can construct models of the plasma environment around the star. These in turn can be used to estimate quantities relevant to SPI.

\section{Outlook}
\label{sec:future}

We expect the emerging trend of claimed radio emission from stellar CMEs, SPI and planetary aurora from LOFAR, FAST, GMRT, JVLA, and ASKAP will become a flood in the next decade, as the expected construction and commissioning of the much more sensitive Square Kilometre Array (SKA; \cite{Dewdney2009}) and ngVLA are completed. Scaling from the sample of LOFAR detected stars would suggest that SKA1-Low ($\sim 50-350$\,MHz) could detect $\sim 10^3$ M~dwarf systems \cite{callingham21,2023A&A...670A.124C}, and SKA1-Mid ($\sim 350$\,MHz -- 15.4\,GHz) and ngVLA (1.2 -- 116\,GHz) will be sensitive even to quiescent radio emission from the nearest stars \cite{Pope2019}. These relatively higher frequencies of SKA1-Mid/ngVLA ($\sim$1-2\,GHz) will likely only probe stellar magnetic fields via SPI, while the lower frequencies of LOFAR/SKA-Low could directly detect auroral emission from Jupiter-mass and larger planets. For tracing aurora from lower mass planets, such as an Earth twin, observations will need to be conducted from space since the Earth's ionosphere reflects emission $\lesssim10$\,MHz back into space. In the long-term, low-frequency exoplanet science will require radio interferometers on the far side of the Moon \cite{Burns2021,Burns2021b}, the best location in the Solar System for avoiding terrestrial radio frequency interference -- provided that lunar satellites and missions do not pollute that pristine location by the time such telescopes can be established. 

The new groundbased radio facilities will need to be matched with both improved theoretical models of stellar plasma environments, and multiwavelength observations that independently constrain stellar activity and exoplanet companions. One of the most important questions that MHD modelling could begin to answer is the extent to which sub-Alfv\'{e}nic SPI affects planetary atmospheric mass loss and habitability. 

Furthermore, more statistical robustness is needed in the field before claiming the detection of radio periodicity from SPI considering the emission mechanism is inherently variable and often dependent on unknown stellar magnetic field properties that can vary in time \cite{bellotti23}. The expected periodicity can also bear a complicated relationship to the orbital and rotational periods \cite{Kavanagh2023}. We suggest the following rule of thumb should be adopted in the field: at least three detections of a stellar system at the same phase, preferably distinct from the rotational phase of the star, should be required before a reliable claim of detection of SPI can be made. Such a standard would prevent premature declaration of the detection of radio emission from SPI, with off-phase observations also vital in ensuring the alignment of detections at a set phase is due to a biased sampling window. Currently, no claim of SPI in the literature has met this standard.

Detecting these variable and periodic signals may benefit from treatment of methods with greater statistical sophistication. For example, using Gaussian Processes \cite{Aigrain2022} or generalized periodograms \cite{VanderPlas2018} could potentially aid in a reliable determination of periodicity from heavily undersampled datasets. A firm attribution of observed radio emission to SPI also depends critically on knowledge of the stellar magnetic field, implying it will be essential to complement SKA facilities with dedicated facilities for contemporaneous spectropolarimetric monitoring of interesting targets. 

As radio emission from SPI may even be biased \textit{against} transiting or edge-on systems \cite{Kavanagh2023}, it will be necessary to continue precise radial velocity (PRV) surveys to determine the existence and orbits of exoplanets inferred from the radio. Red-sensitive and NIR instruments (such as SPIRou \cite{spirou}, the Habitable Planet Finder \cite{hpf}, and CARMENES \cite{carmenes}) will be especially important for probing the population around M~dwarfs where the Alfv\'{en} surface encompasses a large fraction of planetary systems and reaches into habitable zone orbits. As well as stabilized spectrographs for ZDI and PRVs, short-cadence time series photometry is essential for detecting stellar flares. Photometry by itself can determine a stellar flare rate and can help distinguish between SPI and coronal emission \cite{Pope2021}, but simultaneous photometry is ideal since have been observed correlation between some radio bursts and optical flares \cite{Zic2020}.

Considering the significant commitment of observatory resources required to determine radio detections of SPI, it is important the field invests wisely on the most likely candidates. To first order, it is possible to identify the current most likely systems to produce SPI radio emission based on the size of the exoplanet, its proximity to its host star, and the distance of the stellar system from Earth \cite{Griessmeier17PREVIII,2019MNRAS.488..633V}. In this case, the top five candidate systems are 51\,Pegasi, HIP\,65\,A, Tau\,Bo\"{o}tis\,A, 55\,Cancri\,A, and WASP-18 -- many of which have been searched for radio signatures already. 

However, many of these stars likely do not have strong enough magnetic fields to produce ECM emission at $\gtrsim$\,150\,MHz, where our most sensitive telescopes currently operate. Such stars also likely have complicated magnetic field topologies, implying that radio emission may not be consistently beamed to Earth. Therefore, planets close to M~dwarfs may be the best systems to followup considering M dwarfs can posses kilogauss-strong dipolar magnetic fields. GJ\,367, GJ\,436, GJ\,1252, GJ\,3253, GJ\,625, YZ\,Ceti, AU\,Microscopii, and Proxima Centauri could be ideal candidates in such a case -- of which many have been searched \cite{PerezTorres2021,bloot2024}. The main issue with focusing on M~dwarfs for SPI radio signatures is that the star itself is also known to produce radio emission. As discussed above, this implies that a robust detection of SPI will likely involve $\gtrsim$\,200\,hr on radio telescopes since the planets will have a $\sim$1 to 10 day orbital period.

Regardless of the difficulties faced in determining radio emission from SPI, exoplanets, and stars, the potential scientific return is invaluable. Radio observations of stellar systems can provide information about the planet, star, and space weather that are not directly attainable at any other wavelength. Determining the potential habitability of exoplanets will be a focus of astronomy in the coming decades -- and radio astronomy is poised to provide key pieces to this puzzle. 

\backmatter

\bmhead{Acknowledgments}

This project was initiated at the Lorentz Center workshop \textit{Life Around a Radio Star}, held 27 June - 1 July 2022 in Leiden, the Netherlands.

JRC thanks the following graduate students and postdoctoral scholars for providing comments on the manuscript from the perspective of scientists new to the field: Sanne Bloot (ASTRON), Cristina Cordun (ASTRON), Evan Fitzmaurice (Penn. State), David Konijn (ASTRON), Kristo Ment (Penn. State), and Timothy Yiu (ASTRON).  

This research has made use of the NASA Exoplanet Archive, which is operated by the California Institute of Technology, under contract with the National Aeronautics and Space Administration under the Exoplanet Exploration Program.

BJSP acknowledges and pays respect to the traditional owners of the land on which the University of Queensland is situated, and to their Ancestors and descendants, who continue cultural and spiritual connections to Country. He acknowledges funding from the ARC DECRA DE21 scheme and the Big Questions Institute.

RDK acknowledges funding from the Dutch Research Council (NWO) for the e-MAPS (exploring magnetism on the planetary scale) project (project number VI.Vidi.203.093) under the NWO talent scheme Vidi. 

SB acknowledges funding from the Dutch Research Council (NWO) for the `Exo-space weather and contemporaneous signatures of star-planet interactions' of the research programme `Open Competition Domain Science- M' (number OCENW.M.22.215).

MD acknowledges support from the INAF funding scheme Fundamental Research in Astrophysics 2022 (mini grant ``A pilot study to explore the potential of SRT in detecting nearby radio-emitting stars with confirmed or candidate exoplanets, supported by a radial velocity follow-up'').

PZ acknowledges funding from the ERC N$^\circ 101020459-$Exoradio.

SM acknowledges funding from NSF AST-2108512 for a precision NIR M dwarf radial velocity survey with HPF, from NASA XRP investigating radio detected M dwarfs.

JM acknowledges funding from the French National Research Agency (ANR) under contract number ANR-18-CE31-0019 (SPlaSH).

AAV acknowledges funding from the European Research Council (ERC) under the European Union's Horizon 2020 research and innovation programme (grant agreement No 817540, ASTROFLOW).

GS acknowledges support provided by NASA through the NASA Hubble Fellowship grant HST-HF2-51519.001-A awarded by the Space Telescope Science Institute, which is operated by the Association of Universities for Research in Astronomy, Inc., for NASA, under contract NAS5-26555.

BK acknowledge funding from the European Research Council under the European Union’s Horizon 2020 research and innovation programme (grant agreement No 865624,
GPRV).

JDT was supported for this work by NASA through the NASA Hubble Fellowship grant $\#$HST-HF2-51495.001-A awarded by the Space Telescope Science Institute, which is operated by the Association of Universities for Research in Astronomy, Incorporated, under NASA contract NAS5-26555.

JS received funding from the European Research Council (ERC) under the European Union's Horizon 2020 research and innovation programme (grant agreement No. 884711).

JMG acknowledges support by the ``Programme National de Plan\'{e}tologie'' (PNP) of
CNRS/INSU co-funded by CNES and by the ``Programme National de Physique
Stellaire'' (PNPS) of CNRS/INSU co-funded by CEA and CNES.

MMK acknowledges support from the Heising-Simons Foundation through the 51 Pegasi b Fellowship grant 2021-2943.

\section*{Declarations}

\begin{itemize}
\item This project was partly funded by the Lorentz Centre at Leiden University.
\item Conflict of interest: The authors do not have any conflicts of interest to report.
% \item Ethics approval 
% \item Consent to participate
% \item Consent for publication
% \item Availability of data and materials
\item Authors' contributions: Callingham organized overall structure, acted as chief editor, led the replies to the referees, and edited all contributions into a cohesive text with Pope and Kavanagh. Kavanagh produced Figure\,\ref{fig:sketch}. Callingham, Nichols, Rigney, Saur, Turner, and Zarka were the principal contributors to the section on Jupiter and the Sun; Daley-Yates, G\"{u}del, G\"{u}nther, Osten, Pope, and Villadsen, were the principal contributors to the stellar flares and CMEs section; Callingham, Kavanagh, Perez-Torres, Saur, Vedantham, Vidotto, and Zarka were the principal contributors to the SPI section; Grie\ss meier and Turner about radio emission from planet aurorae. All authors reviewed the final text.
\end{itemize}

\bibliographystyle{sn-nature}
\bibliography{bib_natastro_review}% common bib file

\end{document}